\documentstyle[aps,preprint,12pt]{revtex}

\begin{document}

\newcommand{\bel}{\begin{equation}\label}
\newcommand{\f}{\frac}
\newcommand{\bee}{\begin{equation}}
\newcommand{\ee}{\end{equation}}
\newcommand{\br}{\begin{eqnarray}}
\newcommand{\brr}{\begin{eqnarray*}}
\newcommand{\er}{\end{eqnarray}}
\newcommand{\err}{\end{eqnarray*}}
\newcommand{\pr}{\partial}
\newcommand{\th}{\theta}
\newcommand{\non}{\nonumber \\}

\renewcommand{\thesection}{\arabic{section}}
\renewcommand{\theequation}{\thesection.\arabic{equation}}

\title{
\hfill\parbox{4cm}
{\normalsize 
IMSc/99/04/16 \\
PRL-TH-99/002 \\
hep-th/yymmxxx}\\
\vspace{1cm}
The Character of the Exceptional Series of Representations of $SU(1,1)$}
\author{Debabrata Basu}
\address{Department of Physics and Meteorology, 
Indian Institute of Technology, \\
Kharagpur - 721 302, West Bengal,
India.\\email - dbasu@phy.iitkgp.ernet.in}
\author{Subrata Bal}
\address{Institute of Mathematical Sciences, 
C. I. T. Campus, Chennai - 600 113, India. \\email - subrata@imsc.ernet.in}
\author{K. V. Shajesh}
\address{Physical Research Laboratory, Navrangpura, 
Ahmedabad - 380 009, India. \\ email - kvs@prl.ernet.in}
\maketitle

\newpage
\begin{abstract}

The character of the exceptional series of representations of $SU(1,1)$
is determined by using Bargmann's realization of the
representation in the Hilbert space $H_\sigma$ of functions defined on
the unit circle. The construction of the integral kernel of the group ring
turns out to be especially involved because of the non-local metric
appearing in the scalar product with respect to which the representations
are unitary. Since the non-local metric disappears in the `momentum space'
$i.e.$ in the space of the Fourier coefficients the integral kernel is
constructed in the momentum space, which is transformed back to yield the
integral kernel of the group ring in $H_\sigma$. The rest of the
procedure is parallel to that for the principal series treated in a
previous paper. The main 
advantage of this method is that the entire
analysis can be carried  out within  the canonical framework of
Bargmann.
\end{abstract}

\newpage
\section{Introduction}\label{int}
\setcounter{equation}{0}

It is well known that the traditional definition of character breaks down for 
infinite dimensional representations of locally compact groups. For such 
representations character was defined by Gel'fand $\!\!^1$ 
and coworkers as

\[
Tr(T_{x})=\int K(z,z)d\lambda(z)
\]

\noindent
where $K(z,z_1) $ is the integral kernel of the group ring 

\[
T_{x}f(z)=\int K(z,z_{1})f(z_{1})d\lambda(z_{1}).
\]

\noindent
The operator $T_x $ of the group ring is defined by 

\[ T_{x}=\int d\mu(g)
     x(g)T_{g}
\]

\noindent
where $x(g) $ is a test function on the group which vanishes outside a 
bounded set, $d\mu(g) $ is the left and right invariant measure (assumed
coincident) on the group and $ g \rightarrow T_g $ is 
a unitary representation of the group realized in the Hilbert
space $H$ of functions $f(z)$ with the scalar product 

\[(f,g)=\int \overline{f(z)}g(z) d\lambda(z).\]

\noindent
It was shown by Gel'fand $\!\!^1$ and coworkers that $Tr(T_x)$ can be 
written in the form 

\begin{equation}\label{iDtrtx} Tr(T_{x})=\int
     x(g)\pi(g)d\mu(g).
\end{equation}

\noindent
The function $\pi(g)$ is the character of the
representation $g \rightarrow T_{g}$.

In a previous paper $\!\!^2$ (I) 
the character problems of $SU(2)$ and $SU(1,1)$ 
were reexamined  from the standpoint of a physicist by employing the 
powerful Hilbert space method of Bargmann $\!\!^3$ 
and Segal $\!\!^4$ which
was shown to yield a completely unified treatment for $SU(2)$ and the 
discrete series of representations of $SU(1,1)$. The main advantage of 
this method is that the entire analysis can be carried out within the 
canonical framework of Bargmann $\!\!^5$. The representations of the 
positive discrete series were realized in I in the Hilbert space of 
functions analytic within the open unit disc. For the principal series the
carrier space was chosen to be the Hilbert space of functions
defined on the unit circle. 

It is the object of this paper to extend the method of I to the exceptional 
or the supplementary series of representations of $SU(1,1) $.
This representation is realized, as for the principal series, in the space 
of functions defined on the unit circle. The construction of the integral 
kernel of the group ring is, however, more involved for the exceptional 
representations because the scalar product contains a 
nonlocal metric $\!\!^{5,6}$

\bee
(f, g) =
\f{c}{4 \pi^2}
\int_0^{2\pi} d \theta_1
\int_0^{2\pi} d \theta_2
~\overline{f(\theta_1)} 
~e^{i(\f{1}{2} - \sigma)(\theta_2 - \theta_1)}
\left[ 1 - e^{i(\theta_2 - \theta_1)} \right]^{(2 \sigma - 1)}
g(\theta_2)
\ee

\noindent
where  
\[
c = \f{\pi~ 2^{(1-2 \sigma)} e^{i\pi
(\sigma - \f{1}{2})} }{ B ( \f{1}{2} , \sigma)}.
\]

\noindent
This difficulty is resolved by expanding the functions in Fourier series 

\[ f(\theta) = \f{1}{\sqrt{2\pi}}
\sum_{m = - \infty}^{\infty} a_m e^{im\theta}.
\]

\noindent
Since the nonlocal metric disappears in the `momentum space' 
$i.e.$ in the space of the Fourier coefficients $a_m$ 
the integral kernel is constructed in the 
momentum space, which is transformed back to yield 
the integral kernel of the group ring

\bee
T_x f(\theta) =
\f{c}{4 \pi^2} \int \int Q(\theta, \theta_1) 
e^{i(\th_2 - \th_1)(\f{1}{2} - \sigma)}
\left[ 1 - e^{i(\th_2 - \th_1)} \right]^{(2 \sigma - 1)}
f(\theta_2) d \theta_1 d\theta_2
\ee

\noindent
in the space of the functions $f(\th)$. The rest of the procedure is parallel 
to that for the principal series of representations as outlined in I. An 
important common feature of the principal and supplementary 
series is that the elliptic elements of $SU(1,1)$ 
do not contribute to their character.

\section{The exceptional representations of the group $SU(1,1)$}
\setcounter{equation}{0}

To make this paper self-contained we start with the basic properties of the group
$SU(1,1)$ which consists of pseudounitary unimodular matrices

\begin{equation}\label{udUD}
     u=\left(\begin{array}{cc}
        \alpha &\beta\\
        \bar{\beta} &\bar{\alpha}
     \end{array}\right),\hspace{.5 in}
|\alpha|^{2}-|\beta|^{2}=1
\end{equation}
and is isomorphic to the group SL(2,R) of real unimodular matrices,
\begin{equation}\label{udHslD}
     g= \left(\begin{array}{cc} 
	a &b\\c &d 
     \end{array} \right), \hspace{.5 in}
ad-bc=1.
\end{equation}
A particular choice of the isomorphism kernel is 
\begin{equation}\label{etamat}
     \eta=\frac{1}{\sqrt{2}}\left( \begin{array}{cc}
     1 &i\\ i &1 \end{array}\right)
\end{equation}
so that 
\[
  u=\eta g \eta^{-1}
\]
\begin{equation}\label{etaalbeta}
     \alpha=\frac{1}{2}\left[(a+d)-i(b-c)\right],\hspace{.5 in}
     \beta= \frac{1}{2}\left[(b+c)-i(a-d)\right].
\end{equation}
The elements of the group SU(1,1) may be divided into three subsets : 
(a)elliptic, (b)hyperbolic and (c)parabolic. We define them as follows. 
Let $\alpha=\alpha_{1}+i\alpha_{2}$, $\beta=\beta_{1}+i\beta_{2}$
so that
\[\alpha_{1}^{2}+\alpha_{2}^{2}-\beta_{1}^{2}-\beta_{2}^{2}=1.\]
The elliptic elements are those for which 
\[\alpha_{2}^{2}-\beta_{1}^{2}-\beta_{2}^{2}>0.\]
Hence if we set \[
\alpha'_{2}=\sqrt{\alpha_{2}^{2}-\beta_{1}^{2}-\beta_{2}^{2}} \]
we have
\[ \alpha_{1}^{2}+\alpha_{2}^{\prime^{2}}=1 \]
 so that $-1<\alpha_{1}<1$.  

     On the other hand the hyperbolic elements of SU(1,1) are those for 
which
\[\alpha_{2}^{2}-\beta_{1}^{2}-\beta_{2}^{2}<0.\]
Hence if we write
\[ \alpha'_{2}=\sqrt{\beta_{1}^{2}+\beta_{2}^{2}-\alpha_{2}^{2}} \]
we have 
\begin{equation}\label{uhyp}
\alpha_{1}^{2}-\alpha_{2}^{\prime^{2}}=1
\end{equation}
 so that $|\alpha_{1}|>1$.

     We exclude the parabolic class corresponding to
\[\alpha_{2}=\sqrt{\beta_{1}^{2}+\beta_{2}^{2}}\]
as this is a submanifold of lower dimensions.

     If we diagonalize the SU(1,1) matrix (~\ref{udUD}), the eigenvalues
are given by 
\[\lambda=\alpha_{1} \pm \sqrt{\alpha_{1}^{2}-1}\]

\noindent
For the elliptic elements $~\alpha_{1}=\cos(\frac{\theta_{0}}{2})$, 
$0<\theta_{0}<2\pi$
so that $\lambda = \exp( \pm i\frac{\theta_{0}}{2})$. 
For the hyperbolic elements $|\alpha_1| > 1$ so that setting 
$\alpha_1 = \epsilon \cosh \f{t}{2}$, $\epsilon = sgn \lambda$
we obtain the eigenvalues as 
$\epsilon \exp ( \pm \epsilon \f{t}{2})$. 
Since the diagonal matrix

\begin{equation}\label{bst}
     \epsilon(t) =\left(\begin{array}{cc}
     sgn \lambda~ e^{sgn \lambda (\frac{t }{2})} &0\\
     0 &sgn \lambda~ e^{-sgn \lambda (\frac{t }{2})}\end{array}\right)
\end{equation}
 belongs to SL(2,R), it can be regarded as the diagonal form of the
 matrix $g$ given by equations(~\ref{udHslD}) and (~\ref{etaalbeta}) with
$|\alpha_{1}|=\frac{|a+d|}{2}$. 

\noindent
Following Bargmann we realize the representations of the 
exceptional series in the Hilbert space $H_\sigma$ of functions 
defined on the unit circle. The finite element of the group 
in this realization is given by

\begin{equation}\label{urGRre}
      T_{u}f(z)=|\beta z + \bar{\alpha}|^{-2\sigma -1 } f \left(
     \frac{\alpha z + \bar{\beta}}{\beta z + \bar{\alpha}} \right)
\end{equation}

\noindent
where  

\bee 
- \f{1}{2} < \sigma < \f{1}{2};
\hspace{.5 in}
z = e^{i \theta}, 
\hspace{.2 in}
0 < \theta < 2 \pi. 
\nonumber
\ee

\noindent
This representation is unitary with respect to  the scalar product 

\bel{uscpdt}
(f_1, f_2) =
\f{c}{4 \pi^2}
\int_0^{2\pi} d \theta_1
\int_0^{2\pi} d \theta_2
\overline{f_1(z_1)} 
\left( \f{z_2}{z_1} \right)^{(\f{1}{2} - \sigma)}
\left( 1 - \f{z_2}{z_1} \right)^{(2 \sigma - 1)}
f_2(z_2)
\ee

\noindent
where, 

\bel{const}  
z_k = e^{i \theta_k}, 
\hspace{.2 in}
k = 1,2;  
\hspace{.5 in}
c = \f{\pi~ 2^{(1-2 \sigma)} e^{i\pi
(\sigma - \f{1}{2})} }{ B ( \f{1}{2} , \sigma)}.
\ee

\noindent
The integral converges in the usual sense for $0< \sigma < \f{1}{2}$. 
For $-\f{1}{2} < \sigma < 0 $ the integral is to be understood in the 
sense of its regularization $\!\!^7$.

Setting $z = - i e^{i \th}$ the finite group element takes the form,

\bee
 T_{u}f(- i e^{i \th})=|-i\beta   e^{i \th} + 
	\bar{\alpha}|^{-2\sigma -1 } f \left(
     \frac{-i \alpha   e^{i \th} + 
	\bar{\beta}}{-i \beta   e^{i \th} + \bar{\alpha}} \right).
\ee

\noindent
We now introduce as usual the operator of the group ring

\bee
T_x = \int d \mu(u) x(u) T_u
\ee

\noindent
where $x(u)$ is an arbitrary test function on the group which 
vanishes outside a bounded set and $d\mu(u)$ is the left and 
right invariant measure on the
group. The action of the operator $T_x$ is given by 

\bee                                                                          
 T_{x}f(- i e^{i \th})=\int d\mu(u) x(u) 
\left|-i\beta   e^{i \f{\th}{2}} + 
	\bar{\alpha} e^{-i \f{\th}{2}} \right|^{-2\sigma -1 } 
f \left(\frac{- i\alpha   e^{i \f{\th}{2}} + 
	\bar{\beta} e^{-i \f{\th}{2}}}
{-i \beta   e^{i \f{\th}{2}} + 
	\bar{\alpha} e^{-i \f{\th}{2}}} \right).
\ee

\noindent
We now make a left translation

\[
u \rightarrow \underline{\theta} ^{-1} ~u
\]
where
\[
\underline{\theta} = \left(\begin{array}{cc}
                         e^{i \theta /2} & 0 \\
                         0 & e^{-i \theta /2}   
                     \end{array}\right).
\]

\noindent 
We, therefore, obtain

\bee
 T_x  f(-ie^{i \theta})     
 =  
 \int d \mu (u) x(\underline{\theta} ^{-1} u) 
         ~ {\mid -i\beta  + \bar{\alpha}
      \mid }^{-2 \sigma - 1}  f\left( \f{-i\alpha  + \bar{\beta}}
               {-i\beta  + \bar{\alpha}}\right). 
\ee

\noindent
We now map the $SU(1,1)$ matrix $u$ onto the $SL(2,R)$
matrix $g$ by using the isomorphism kernel $\eta$ given 
by (\ref{etamat}) and (\ref{etaalbeta}) and perform the Iwasawa 
decomposition 

\bel{3.9a}
g = k ~\theta _2
\ee
where
\bel{3.9b}
k = \left(\begin{array}{cc}
         k_{11} & k_{12} \\
         0 & k_{22}
     \end{array}\right),
\hspace{.5in}
k_{11} k_{22} = 1
\ee
belongs to the subgroup $K$ of real triangular matrices
of determinant unity and $\theta _2 \in \Theta$ where $\Theta$  is the
subgroup of pure rotation matrices
\bel{3.9c}
\theta _2 = \left(\begin{array}{cc}
             \cos (\theta _2 /2) & - \sin (\theta _2 /2) \\
             \sin (\theta _2 /2) & \cos (\theta _2 /2)
            \end{array}\right).
\ee

\noindent
As in I we now introduce the following convention. The letters 
without a bar below it will denote the $SL(2,R)$
matrices or its subgroups and those with a bar below it
will denote their $SU(1,1)$ image. For instance
\[
\underline{k} = \eta ~k~ \eta ^{-1} = \f{1}{2}
        \left(\begin{array}{cc}
         k_{11} + k_{22} - ik_{12}~~~~~ &
                     k_{12} - i(k_{11} - k_{22}) \\
         k_{12} + i (k_{11} - k_{22}) ~~~~~&
                     k_{11} + k_{22} + i k_{12}
        \end{array}\right)
\]
\[
\underline{\theta} _2 = 
        \left(\begin{array}{cc}
         e^{i \theta _2 /2} & 0 \\
         0 & e^ {-i \theta _2 /2}
        \end{array}\right).
\]
Thus the decomposition (\ref{3.9a}) can also be written as
\bel{3.11}
u = \underline{k}~ \underline{\theta} _2
\ee
which yields
\[
-i ~\alpha ~+~ \bar{\beta} ~=~ -i ~k_{22} ~e^{i \theta _2 /2}; 
\hspace{.5in}
-i ~\beta ~+~ \bar{\alpha} ~=~ k_{22} ~e^{-i \theta _2 /2}.
\]

\noindent
Hence setting $f(-ie^{i \theta}) = g(\theta)$ 
we obtain,
\bel{3.13}
T_x ~g(\theta) = \int 
     x(\underline{\theta} ^{-1} ~\underline{k}
            ~\underline{\theta} _2 )
    ~ {\mid k_{22} \mid }^{-2 \sigma - 1} ~g (\theta _2)
     ~d \mu (u).
\ee

\noindent 
It can be shown that under the decomposition
(\ref{3.9a}) or equivalently (\ref{3.11}) the invariant
measure decomposes as 
\bel{3.14}
d \mu (u) = \f{1}{2} ~d \mu _l (g) = 
  \f{1}{2} ~d \mu _r (g) =
     \f{1}{4} ~d \mu _l (k) ~d \theta _2.
\ee

\noindent
Substituting the decomposition (\ref{3.14}) in eqn.
(\ref{3.13}) we have

\bel{txgth}
T_x ~g(\theta) = \int 
     K(\theta ,\theta _2) ~g(\theta _2)
       ~d \theta _2 
\ee
where
\bel{suplker1}
K(\theta ,\theta _2) = \f{1}{4} ~\int
   x(\underline{\theta} ^{-1} ~\underline{k}
       ~\underline{\theta} _2 )
   ~{\mid k_{22} \mid}^{-2 \sigma - 1} 
      ~d \mu _l (k).
\ee

\noindent
It must be pointed out that eqn. (\ref{txgth}) does not yield the integral 
kernel of the operator $T_x$ of the group ring because $T_x$ now is an 
operator in the space $H_\sigma$ in which the scalar product is given
by eqn.(\ref{uscpdt}). It is, therefore, not clear a priori that 

\bel{trace}
\int K(\th, \th) d \th
\ee

\noindent
is the trace of the operator $T_x$. Nevertheless, 
we shall show that (\ref{trace})
is the trace of the operator $T_x$.

To write the action of $T_x$ on $g(\th)$ in the form consistent with the scalar
product (\ref{uscpdt}) we pass over to the `momentum space' 
(space of the Fourier coefficients)
\bee
g(\th) = \f{1}{\sqrt{2\pi}} \sum_{m = -\infty}^{\infty}
a_m e^{im \th}
\ee

\noindent 
where

\bel{amdef}
a_m =  \f{1}{\sqrt{2\pi}} \int_0^{2\pi} g(\th)e^{-im \th} d \theta.
\ee

\noindent
We then obtain

\bel{uscpdtf1}
(g_1, g_2) = \sum_{m = - \infty}^{\infty} \overline{a}_m b_m \rho_m
\ee

\noindent
where $b_m$ is the Fourier coefficient of $g_2(\th)$, 

\bee
b_m = \f{1}{\sqrt{2 \pi}} \int_0^{2\pi} g_2(\th)e^{-im \th} d \th
\ee

\noindent 
and

\bel{rhom}
\rho_m = \f{1}{2\pi} \f{\Gamma(\f{1}{2} + \sigma) \Gamma(\f{1}{2} - \sigma +m)}
{\Gamma(\f{1}{2} - \sigma) \Gamma(\f{1}{2} + \sigma +m)}.
\ee

\noindent 
Hence

\bee
|| g||^2 = \sum_{m= - \infty}^\infty |a_m|^2 \rho_m 
\ee

\noindent
We can, therefore, define the scalar product in the momentum space as

\bee
(a,b) = \sum \overline{a}_m b_m \rho_m 
\ee

\noindent
where $\rho_m$ as given by eqn.(\ref{rhom}) is a positive metric.

The operator of the group ring in the momentum space is given by

\bee
T_x a_m = \f{1}{2 \pi} \sum_{n = -\infty}^\infty a_n \int_0^{2\pi}
\int_0^{2 \pi} K(\th, \th_2) e^{-i(m \th - n \th_2)} d \th d \th_2.
\ee

\noindent
If we now expand $K(\th, \th_2)$ in a Fourier series,

\bee
K(\th, \th_2) = \f{1}{2\pi} \sum L_{mn} e^{i(m \th - n \th_2)}
\ee

\noindent
we have 

\bel{txam}
T_x a_m =  \sum_{n = -\infty}^\infty L_{mn} a_n 
\ee

\noindent
where

\bee
L_{mn}  = \f{1}{2\pi}  \int_0^{2\pi}  \int_0^{2\pi} K(\th, 
\th_2) e^{-i(m \th - n \th_2)} d \th d \th_2. 
\ee

\noindent
We now define

\bel{modd}
L_{mn} = \Gamma_{mn} \rho_n.
\ee

\noindent
Thus

\bee
T_x a_m = \sum_n \Gamma_{mn} \rho_n a_n.
\ee

\noindent
We shall now show that 

\bel{rho1}
\rho_m = \f{c}{4 \pi^2} \int_0^{2\pi} e^{i(\f{1}{2} - \sigma +m)\alpha}
(1 - e^{i\alpha})^{2 \sigma -1} d \alpha.
\ee

\noindent
Setting $z = e^{i \alpha}$ the integral on the r.h.s. can be recast as 
an integral over the unit circle $S$. Since the only singularities of the 
subsequent integrand are the branch points at $z = 0 $ 
and $z = 1$ the unit circle
$S$ can be deformed to a contour $\sum$ that starts from 
$z =1$ along the positive
real axis, encircles the point $z = 0 $ once counterclockwise 
and returns to the 
point $z = 1$ along the positive real axis. The integral is, 
therefore, the contour
integral representation of the beta function regularized at the origin,

\bel{conint}
\int_1^{0_+} t^{(\f{1}{2} - \sigma +m -1)} (1 - t)^{2\sigma -1} dt =
\left[ e^{2\pi i (\f{1}{2} - \sigma )} - 1 \right] 
B \left( \f{1}{2} - \sigma +m,
2 \sigma \right).
\ee

\noindent
The eqn. (\ref{conint}) in conjunction with eqn. (\ref{const}) 
immediately yields
eqn. (\ref{rho1}). 

We now pass over from the momentum space to the space of functions $g(\th)$:

\bel{txinth}
T_x g(\th) = \f{1}{\sqrt{2\pi}} \sum_{m = -\infty}^\infty \sum_{n = -\infty}^\infty 
e^{i m \th} \Gamma_{mn} \rho_n a_n.
\ee

\noindent
We now substitute eqn.(\ref{rho1}) and (\ref{amdef}) in eqn.(\ref{txinth}).
Thus 

\bee
T_x g(\th) = \f{c}{8\pi^3} \int_0^{2\pi} \int_0^{2\pi}  
	d \alpha d \th_2 g(\th_2)
e^{i(\f{1}{2} - \sigma) \alpha } (1 - e^{i\alpha})^{2 \sigma -1}
\sum \f{L_{mn}}{\rho_n} e^{i[m\th -n(\th_2 - \alpha)]}.
\ee

\noindent
Since $L_{mn}$ is the Fourier coefficient of $K(\th, \th_2) $ and $\rho_n$ 
is given by eqn.(\ref{rhom}) the function

\bel{qthal}
Q(\th, \th_2 - \alpha) = \f{1}{2\pi} \sum  \f{L_{mn}}{\rho_n} e^{i[m\th -n(\th_2 - \alpha)] }
\ee

\noindent
is known and well defined. Hence we have 

\bee  
T_x g(\th) = \f{c}{4\pi^2} \int_0^{2\pi} \int_0^{2\pi}  d \alpha 
d \th_2 g(\th_2) Q(\th, \th_2 - \alpha)
e^{i(\f{1}{2} - \sigma) \alpha } (1 - e^{i\alpha})^{2 \sigma -1}. 
\ee   

\noindent
Finally setting $ \alpha = \th_2 - \th_1$ we have

\bel{txinthf}
T_x g(\th) = \f{c}{4\pi^2} \int_0^{2\pi} \int_0^{2\pi}  d \th_1 
d \th_2 g(\th_2) Q(\th, \th_1 )  
e^{i(\f{1}{2} - \sigma) (\th_2 - \th_1) } 
(1 - e^{i(\th_2 - \th_1)})^{2 \sigma -1} 
\ee  

\noindent
which is in the form consistent with the scalar product  (\ref{uscpdt}).
$Q(\th, \th_1 )$
is, therefore, the integral kernel of the group ring. 
Comparing eqn.(\ref{txinthf})
with eqn.(\ref{txgth}) we have

\bel{kthth2}
K(\th, \th_2) =  \f{c}{4\pi^2} \int_0^{2\pi} 
Q(\th, \th_1 )
e^{i(\f{1}{2} - \sigma) (\th_2 - \th_1) } 
(1 - e^{i(\th_2 - \th_1)})^{2 \sigma -1} d \th_1. 
\ee 

\noindent
From eqn.(\ref{txam}) $Tr(T_x)$ is given by

\bel{trtx1}
Tr(T_x) = \sum_{n = -\infty}^\infty L_{nn}
= \sum_{n = -\infty}^\infty \Gamma_{nn} \rho_n.
\ee

\noindent
Substituting the integral representation (\ref{rho1}) in eqn.(\ref{trtx1})
we obtain

\bel{trtx2}
Tr(T_x) = \f{c}{4\pi^2} \int d \alpha 
~e^{i(\f{1}{2} - \sigma) \alpha } (1 - e^{i\alpha})^{2 \sigma -1}
\sum_n \Gamma_{nn} e^{i n \alpha}.
\ee

\noindent
Now from eqns.(\ref{modd}) and (\ref{qthal}) 

\bee
\Gamma_{nn} = \f{1}{2\pi} \int Q(\th, \th_1) 
e^{- i n (\th - \th_1)} d\th d \th_1.
\ee

\noindent
The above equation in conjunction with eqn.(\ref{trtx2}) yields,

\bel{trtx3}  
Tr(T_x) = 
\frac{c}{4 \pi^2}
\int \int d \th d \th_1 Q(\th, \th_1) 
\int_0^{2\pi}  d \alpha 
~e^{i(\f{1}{2} - \sigma) \alpha } (1 - e^{i\alpha})^{2 \sigma -1}
\f{1}{2\pi}\sum_{n = -\infty}^\infty
e^{i n (\alpha - \th + \th_1)}.
\ee

\noindent
The summation over $n$ appearing on the r.h.s. of eqn.(\ref{trtx3}) can
be carried out using the formula 

\bel{deltaf}
\f{1}{2\pi}\sum_{n = -\infty}^\infty e^{i n \beta} = \delta(\beta).
\ee

\noindent
Eqn.(\ref{deltaf}) immediately leads to 

\bel{trtx4}
Tr(T_x) = 
\frac{c}{4 \pi^2}
\int  Q(\th, \th_1) e^{i(\f{1}{2} - \sigma) (\th - \th_1)}
[ (1 - e^{i(\th-\th_1)})^{2 \sigma -1}
d \th d \th_1.
\ee

\noindent
Finally using eqn.(\ref{kthth2}) we have
\begin{eqnarray}\label{3.16}
Tr~(T_x) 
&=& \int K(\th,\th) d \th 
\nonumber \\
&=& \frac{1}{4} ~\int 
   x(\underline{\theta} ^{-1} ~\underline{k}
         ~\underline{\theta}) 
   ~{\mid k_{22} \mid }^{-2\sigma -1} 
     ~d \mu _l (k) ~d \theta.
\end{eqnarray}

\noindent
The rest of the calculation is parallel to that of the principal series
outlined in I. However, to make the paper self-contained we reproduce the 
steps here. 

We assert that 

\[
u =
\underline{\theta} ^{-1} ~\underline{k} ~\underline{\theta}
\]

\noindent 
represents a hyperbolic element of $SU(1,1)$ because the equation

\begin{equation}\label{3.17b}
2 \alpha_1 = k_{11} + k_{22} = \f{1}{k_{22}}  + k_{22} 
\end{equation}

\noindent 
has no real solution for $\alpha_1 = \cos \f{\th_0}{2} $. Thus the
elliptic elements of $SU(1,1)$ do not contribute to the character of the
exceptional series of representations.

We shall now show that every hyperbolic
element of $SU(1,1)$ ( i.e those with \\ 
$\mid\alpha_1\mid=\frac{\mid(a+d)\mid}{2}>1$
) can be
represented as
\begin{equation}\label{3.18a}
u = 
\underline{\theta} ^{-1} ~\underline{k} ~\underline{\theta}
\end{equation}
or equivalently as 
\begin{equation}\label{3.18b}
g = \theta ^{-1} ~k ~\theta.
\end{equation}
Here $k_{11} = \lambda ^{-1}$, $k_{22} = \lambda$
are the eigenvalues of the matrix $g$ taken in 
any order.

    We recall that every $g \in SL(2,R)$ for the 
hyperbolic case can be diagonalized as
\[
v' ~g ~v'^{-1} = \delta
\]
where
\[
\delta = \left(\begin{array}{cc}
          \delta _1 & 0 \\
          0 & \delta _2
         \end{array}\right),
\hspace{.5in}
\delta _1 \delta _2 = 1,
\hspace{.5in}
\delta_1, \delta_2 ~~real, 
\]
belongs to the subgroup $D$ of real diagonal
matrices of determinant unity and \\
$v'\in SL(2,R)$. If we write the Iwasawa decomposition
for $v'$,
\[
v' = k' ~\theta
\]
then 
\[
g = \theta ^{-1} ~{k'}^{-1} ~\delta ~k' ~\theta.
\]
Now ${k'}^{-1} ~\delta ~k' \in K$ so that writing
$k = {k'} ^{-1} ~\delta ~k'$ we have the decomposition
(\ref{3.18b}) in which
\[
k_{11} = \delta _1 = \lambda ^{-1} ,~~~~
    k_{22} = \delta _2 = \lambda.
\]

\noindent
    If these eigenvalues are distinct then for
a given ordering of them the matrices $k$,
$\theta$ are determined uniquely by the 
matrix $g$. 
It follows that for a given choice of $\lambda$
the parameters $\theta$ and $k_{12}$ are
uniquely determined. We note that there are 
exactly two representations of the matrix $g$
by means of formula (\ref{3.18b}) corresponding
to two distinct possibilities
\[
k_{11} = sgn \lambda {\mid \lambda \mid}^{-1} =
  sgn \lambda ~e^{t/2},
\hspace{.5in}
k_{22} = sgn \lambda ~\mid \lambda \mid
     = sgn \lambda ~e^{- t/2}
\]
\[
k_{11} = sgn \lambda \mid \lambda \mid =
  sgn \lambda ~e^{- t/2},
\hspace{.5in}
k_{22} = sgn \lambda ~\mid \lambda \mid^{-1} 
     = sgn \lambda ~e^{t/2}.
\]
Let us now remove from $K$ the elements with
$k_{11} = k_{22} =1$. This operation cuts 
the group $K$ into two connected disjoint
components. Neither of these components 
contains two matrices which differ only by
permutation of the two diagonal elements.
In correspondence with this partition the 
integral over $K$ is
represented in the form of a sum of two
integrals,
\begin{equation}\label{3.24}
Tr~(T_x) =\frac{1}{4} \int _ {\Theta} d\theta \int _ {K_1}
  d \mu _l (k) ~{\mid k_{22} \mid}^{-2\sigma -1}
  ~x(
\underline{\theta} ^{-1} ~\underline{k} ~\underline{\theta})
 ~+~ \frac{1}{4} \int _{\Theta} d \theta \int _{K_2}
d \mu _l (k) ~{\mid k_{22} \mid }^{-2\sigma -1} ~x(
\underline{\theta} ^{-1} ~\underline{k} ~\underline{\theta}).
\end{equation}
As $\theta$ runs over the subgroup $\Theta$
and $k$ runs over the components $K_1$ or
$K_2$ the matrix 
\\ $u =  
\underline{\theta} ^{-1} ~\underline{k} ~\underline{\theta}$
runs over the hyperbolic elements of the
group $SU(1,1)$. We shall now prove that in $K_1$
or $K_2$
\begin{equation}\label{3.25}
d \mu _l (k) ~d \theta = \frac{4 \mid k_{22} \mid }
   {\mid k_{22} - k_{11} \mid} ~d \mu (u).
\end{equation}
To prove this we start from the left invariant
differential 
\[
dw = g^{-1} ~dg
\]
where $g \in SL(2,R)$ and $dg$ is the matrix of the 
differentials $dg_{pq}$. i.e. 
\[
dg = \left(\begin{array}{cc}
       da & db \\
       dc & dd 
     \end{array}\right).
\]
The elements $dw$ are invariant under the left
translation $g \rightarrow g_0 g$. Hence
choosing a basis in the set of all $dg$ we 
immediately obtain a  left invariant
measure. For instance choosing $dw_{12}$, 
$dw_{21}$ and $dw_{22}$ as independent invariant
differentials we arrive at the left invariant   
measure on $SL(2,R)$,
\begin{equation}\label{3.27}
d \mu _l (g) = dw_{12} dw_{21}  dw_{22}.
\end{equation}

\noindent
We now write the
decomposition (\ref{3.18b}) as
\[
\theta ~g = k ~\theta
\]

\noindent
which yields 

\begin{equation}\label{3.31}
dw = g^{-1} dg = \theta ^{-1} ~d \mu ~\theta
\end{equation}
where
\begin{equation}\label{3.32}
d \mu = k^{-1} dk ~+~ d \theta ~\theta ^{-1} ~-~
	k^{-1} d \theta ~\theta ^{-1} k.
\end{equation}
In accordance with the choice of the independent
elements of $dw$ as mentioned above we choose
the independent elements of $d \mu$ as $d \mu _{12}$,
$d \mu _{21}$ and $d \mu _{22}$. Eqn.(\ref{3.31}) then
leads to
\begin{equation}\label{3.33a}
dw_{11} + dw_{22} = d \mu _{11} + d \mu _{22}
\end{equation}
\begin{equation}\label{3.33b}
dw_{11} dw_{22} - dw_{12} dw_{21} = d\mu_{11} d\mu_{22}
      - d\mu_{12} d\mu_{21}.
\end{equation}
Further since $Tr(d \mu) = Tr(dw) =0$ we 
obtain from eqns.(\ref{3.33a}) and (\ref{3.33b})
\[
dw_{22}^2 + dw_{12} dw_{21} = d\mu_{22}^2 + d\mu_{12}
     d\mu_{21}
\]
which can be written in the form
\begin{equation}\label{3.35}
d\eta_1^2 + d\eta_2^2 - d\eta_3^2 = 
   {d\eta'}_1^2 + {d\eta'}_2^2 - {d\eta'}_3^2 
\end{equation}
where
\[
\begin{tabular}{ll}
 $d \eta_1 = (dw_{12} + dw_{21}) /2~~~~~~~~~~~~~~$  & $d \eta '_1 = 
            (d \mu_{12} + d \mu_{21}) /2$  \\
 $d \eta_2 = dw_{22}$  & $ d \eta ' _2 = d \mu _{22}$  \\
 $d \eta_3 = (dw_{12} - dw_{21}) /2$ & $ d \eta ' _3 =
            (d \mu_{12} - d \mu_{21}) /2 $.
\end{tabular}
\]
Eqn.(\ref{3.35}) implies that the set $d \eta$ and the 
set $d \eta '$ are connected by a Lorentz transformation.
Since the volume element $d \eta _1 d \eta _2 d \eta _3 $
is invariant under such a transformation we have
\begin{equation}\label{3.37}
d \eta _1 d \eta _2 d \eta _3 = 
   d \eta ' _1 d \eta ' _2 d \eta ' _3.
\end{equation}
But the l.h.s. of eqn.(\ref{3.37}) is $dw_{12} dw_{21} dw_{22} /2$ 
and the r.h.s. is $d\mu_{12} d\mu_{21} d\mu_{22} /2$. 
Hence using eqn.(\ref{3.27}) we easily obtain
\[
d\mu_l (g) = d\mu_{12} d\mu_{21} d\mu_{22}.
\]
We now write eqn.(\ref{3.32}) in the form
\begin{equation}\label{3.39}
d\mu = du + dv
\end{equation}
where $du = k^{-1} dk$ is the left invariant differential
element on $K$ and
\begin{equation}\label{3.40}
dv = d \theta ~\theta ^{-1} ~-~ k^{-1} d \theta ~\theta ^{-1} k.
\end{equation}
In eqn.(\ref{3.39}) $du$ is a triangular matrix whose
independent nonvanishing elements are chosen to be
$du_{12}$ and $du_{22}$ so that
\[
d\mu _l (k) = du_{12} du_{22}.
\]
On the other hand $dv$ is a $2 \times 2$ matrix having
one independent element which is chosen to be $dv_{21}$.
Since the Jacobian connecting $d \mu _{12} d \mu _{21} d
\mu _{22}$ and $du_{12} du_{22} dv_{21}$ is a triangular
determinant having $1$ along the main diagonal we
obtain
\begin{equation}\label{3.42}
d\mu _l (g) = d\mu_l (k)~dv_{21}.
\end{equation}
It can now be easily verified that each element $k
\in K$ with distinct diagonal elements (which is
indeed the case for $K_1$ or $K_2$) can be represented          
uniquely in the form
\begin{equation}\label{3.43}
k = \zeta ^{-1} ~\delta ~\zeta
\end{equation}
where $\delta$ belongs to the subgroup of real
diagonal matrices with unit determinant and $\zeta \in Z$
where Z is a subgroup of $K$ consisting of matrices
of the form
\[
\zeta = \left(\begin{array}{cc}
          1 & \zeta _{12} \\
          0 & 1 
        \end{array}\right).
\]
Writing eqn.(\ref{3.43}) in the form $\zeta ~k = \delta ~\zeta$ we
obtain
\begin{equation}\label{3.45}
k_{pp} = \delta _p~,
\hspace{.5in}
\zeta _{12} = \frac{k_{12}}{(\delta _1 - \delta _2 )}.
\end{equation}
Using the decomposition (\ref{3.43}) we can now write eqn.(\ref{3.40})
in the form
\[
dv = \zeta ^{-1} ~dp ~\zeta
\]
where
\[
dp = d \lambda ~-~ \delta ^{-1} ~d \lambda ~\delta
\]
\[
d \lambda = \zeta ~d \sigma ~\zeta ^{-1}~,
\hspace{.5in}
d \sigma = d \theta ~\theta ^{-1}.
\]
From the above equations it now easily follows
\begin{equation}\label{3.47}
dv_{21} = \frac{\mid \delta _2 - \delta _1 \mid}
   {\mid \delta _2 \mid} ~\frac{d \theta}{2}.
\end{equation}
Substituting eqn.(\ref{3.47}) in eqn.(\ref{3.42}) and using eqns.(\ref{3.14})
and (\ref{3.45}) we immediately obtain eqn.(\ref{3.25}).

       Now recalling that in $K_1$, $\mid k_{22} \mid = e^{- t/2}$
and in $K_2$, $\mid k_{22} \mid = e^{t/2}$ eqn.(\ref{3.24})
in conjunction with (\ref{3.25}) yields
\[
Tr~(T_x) = \int x(u) ~\pi (u) ~d \mu (u)
\]
where the character $\pi (u)$ is given by
\begin{eqnarray}
\pi (u) 
&=& \frac{e^{ \sigma t} + e^{- \sigma t}}
     {\mid e^{t/2} - e^{-t/2} \mid}
\nonumber \\
&=& \f{\cosh \sigma t }{|\sinh \f{t}{2}|}
\nonumber
\end{eqnarray}


\begin{references}
$\!\!^1$
I. M. Gel'fand and M. A. Naimark, 
{\em I. M. Gel'fand - Collected Papers}
(Springer-Verlag, Berlin-Heidelberg, 1988), 
Vol I$\!$I, pp. 41, 182. \\  
I. M. Gel'fand, M. I. Graev and I. I. Pyatetskii--Shapiro, 
{\em Representation Theory and Automorphic Functions} 
(W.B. Saunders, Philadelphia, London, Toronto, 1969),
Chap. 2, pp. 199,202. 
\\

$\!\!^2$
S. Bal, K. V. Shajesh and D. Basu,
J. Math. Phys. {\bf 38}, 
3209 (1997).
\\

$\!\!^3$
V. Bargmann, 
Commun. Pure. Appl. Math. {\bf 14}, 
187 (1961); 
{\bf 20}, 1 (1967);
in {\em Analytic Methods in Mathematical Physics}, 
edited by P. Gilbert and R. G. Newton 
(Gordon and Breach, New York, 1970).
\\

$\!\!^4$
I. E. Segal, 
Ill. J. Math. {\bf 6}, 
500 (1962).
\\

$\!\!^5$
V. Bargmann, 
Ann. Math. {\bf 48}, 
568 (1947).
\\

$\!\!^6$
D. Basu,
J. Math. Phys. {\bf 18}, 
7 (1977).
\\

$\!\!^7$
I. M. Gel'fand and G. E. Shilov, 
{\em Generalized Functions}
(Academic, New York, 1964),
Vol.-I, Chap-1.

\end{references}
\end{document}